\documentclass[runningheads]{llncs}
\usepackage[T1]{fontenc}
\usepackage{graphicx,verbatim}

\usepackage{enumitem}  
\usepackage{amssymb}
\usepackage{hyperref}  
\usepackage{xcolor}   
\usepackage{amsmath}  
\usepackage{multirow}  
\usepackage{diagbox}  
\usepackage{booktabs}  
\usepackage{soul}  

\usepackage{comment}
\usepackage{xcolor}

\begin{document}

\title{Parameter-Efficient Fine-Tuning of 3D DDPM for MRI Image Generation Using Tensor Networks}
\titlerunning{TenVOO: PEFT of 3D DDPM for MRI Image Generation Using TNs}

\author{Binghua Li\inst{1,2,3,}\orcidID{0000-0002-2595-4762}, Ziqing Chang\inst{2}, Tong Liang\inst{2}, Chao Li\inst{2,3}, Toshihisa Tanaka\inst{1,3}, Shigeki Aoki\inst{2}, Qibin Zhao\inst{3,1}, Zhe Sun\inst{2,}\thanks{Corresponding author}}
\authorrunning{Binghua Li, Ziqing Chang, Tong Liang et al.}

\institute{
    Tokyo University of Agriculture and Technology, Tokyo, Japan \\
    \and
    Juntendo University, Tokyo, Japan 
    \\ \email{b.li.qr@juntendo.ac.jp, z.sun.kc@juntendo.ac.jp}    
    \and
    RIKEN Center for Advanced Intelligence Project, Tokyo, Japan
}

\maketitle

\begin{abstract}
We address the challenge of parameter-efficient fine-tuning (PEFT) for three-dimensional (3D) U-Net-based denoising diffusion probabilistic models (DDPMs) in magnetic resonance imaging (MRI) image generation. 
Despite its practical significance, research on parameter-efficient representations of 3D convolution operations remains limited.
To bridge this gap, we propose \textbf{Ten}sor \textbf{Vo}lumetric \textbf{O}perator (TenVOO), a novel PEFT method specifically designed for fine-tuning DDPMs with 3D convolutional backbones.
Leveraging tensor network modeling, TenVOO represents 3D convolution kernels with lower-dimensional tensors, effectively capturing complex spatial dependencies during fine-tuning with few parameters.
We evaluate TenVOO on three downstream brain MRI datasets--\emph{ADNI}, \emph{PPMI}, and \emph{BraTS2021}--by fine-tuning a DDPM pretrained on 59,830 T1-weighted brain MRI scans from the UK Biobank. 
Our results demonstrate that TenVOO achieves state-of-the-art performance in multi-scale structural similarity index measure (MS-SSIM), outperforming existing approaches in capturing spatial dependencies while requiring only $0.3\%$ of the trainable parameters of the original model. Our code is available at \url{https://github.com/xiaovhua/tenvoo}.



\end{abstract}
\section{Introduction}\label{sec:intro}
Diffusion models \cite{sohl2015deep}, particularly denoising diffusion probabilistic models (DDPMs)~\cite{ho2020denoising}, have gained attention for generating high-quality and diverse medical images~\cite{ali2022spot,packhauser2023generation}.
Especially in magnetic resonance imaging (MRI) tasks, these models hold great potential for enhancing clinical workflows, including disease prediction, diagnosis, and treatment in clinical practice~\cite{chung2022mr,dorjsembe2024conditional,pinaya2022brain}
By modeling the reverse diffusion process, DDPMs gradually remove noise to reconstruct clear images. 
This training paradigm offers strong robustness and controllability, enabling models to effectively capture complex data distributions. 
However, their large-scale architecture poses significant computational and efficiency challenges.
As a consequence, training a DDPM typically requires high computational costs, extensive data, and resource-intensive optimization, limiting the practical applications. 
Meanwhile, model customization--widely desired in practice by adapting from large models--is constrained by dataset limitations, the need to store multiple parameter variations, and the complexity of designing effective concept descriptors.

To address these challenges, parameter-efficient fine-tuning (PEFT) techniques have achieved significant attention.
In the existing methods, low-rank approaches \cite{chavan2023one,edalati2022krona,hu2022lora,hyeon2021fedpara} introduce trainable low-rank decompositions of model weights, serving as residual backups that can be merged during inference. 
Unlike adapter-based methods \cite{houlsby2019parameter,zaken2021bitfit}, the low-rank modeling enables efficient fine-tuning while preserving inference speed, making it particularly effective for resource-constrained inferences. 
However, representing conventional structures with low-rank modeling poses challenges in capturing intricate spatial relationships (see numerical results in Section~\ref{sec:exp}).
This limitation is particularly critical in MRI image generation, as the data typically exhibit complex spatial dependencies across multiple anatomical structures and resolution scales.

\begin{figure}[t]
    \centering
    \includegraphics[width=0.85\textwidth]{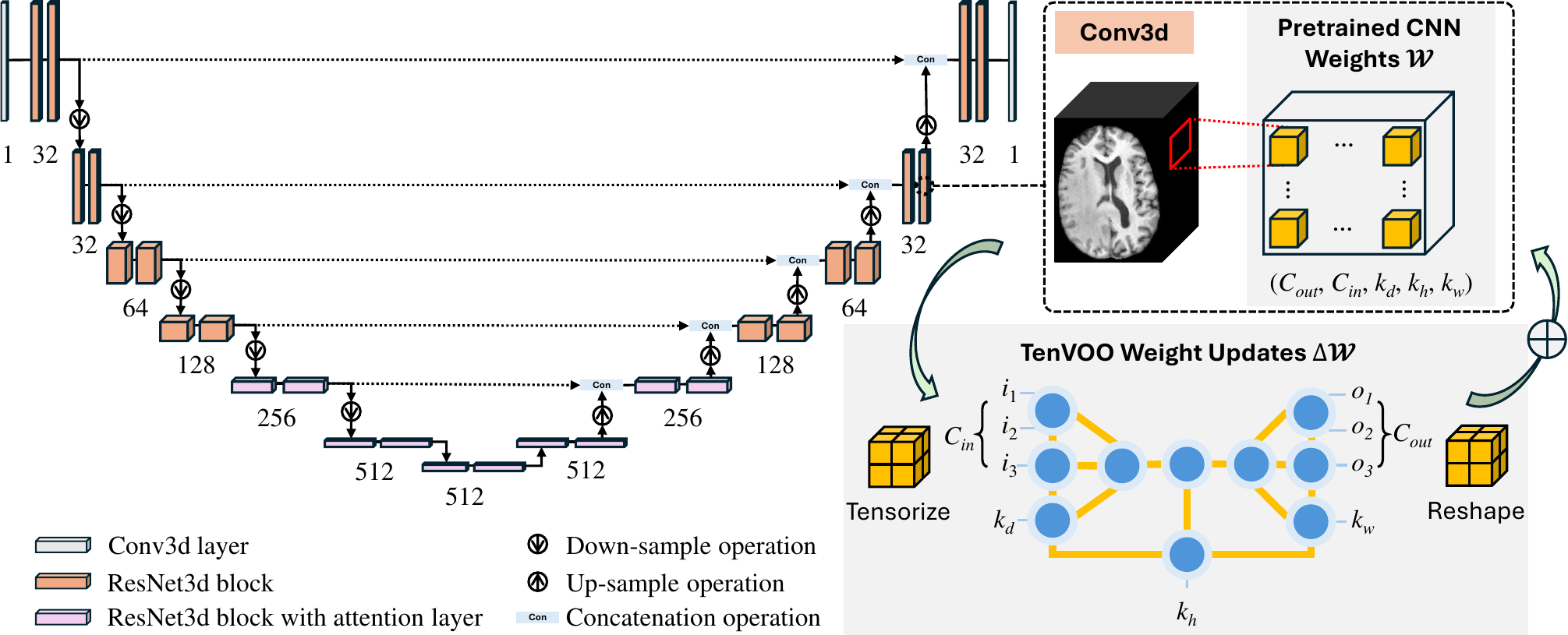}
    \caption{Illustration of the U-Net model used for our DDPM and the TenVOO framework for 3D convolutional layers. TenVOO optimizes only the weight updates $\Delta \mathcal{W}$ via a lightweight tensor network during fine-tuning, while preserving the intricate spatial dependencies of convolution kernels throught tensor contraction.}
    \label{fig1:framework}
\end{figure}

In this work, as shown in Figure \ref{fig1:framework}, we propose a novel PEFT method, \textbf{Tensor} \textbf{Vo}lumetric \textbf{O}perator (TenVOO), for 3D U-Net \cite{ronneberger2015u} fine-tuning, inspired by the concept of tensor networks~\cite{bershatsky2024lotr,chen2024quanta,jie2023fact}. Compared to other low-rank counterparts, TenVOO introduces a trainable yet more compact modeling to represent complex spatial information within 3D convolutional layers, making it a powerful framework to fine-tuning U-Net-based DDPMs. 
Our main contributions are summarized as follows:
\begin{enumerate}[label=(\arabic*)] 
    \item We propose TenVOO, with two variants TenVOO-L and TenVOO-Q, for fine-tuning 3D U-Net-based DDPMs. To the best of our knowledge, TenVOO is the first application of tensor networks for fine-tuning 3D DDPM in MRI image generation.
    \item We conduct experiments on three real-world MRI datasets, including ADNI (Alzheimer’s Disease Neuroimaging Initiative)\cite{jack2008alzheimer}, PPMI (Parkinson’s Progression Markers Initiative)\cite{marek2011parkinson}, and BraTS2021~\cite{baid2021rsna} of brain tumors, to fine-tune a DDPM pretrained on 59,830 T1-weighted brain MRI scans from the UK Biobank.
    The results demonstrate that our method achieves state-of-the-art performance in MRI image generation, effectively capturing complex spatial dependencies while significantly reducing the number of trainable parameters. 
\end{enumerate}

\section{Methodology}\label{sec:method}

\subsection{Preliminaries}
\subsubsection{Denoising diffusion probabilistic models (DDPMs)~\cite{ho2020denoising}.} 
DDPMs are one of the most widely used diffusion models (DMs), designed to reconstruct data by modeling the reverse diffusion process within a probabilistic denoising framework.
Similar to standard DMs \cite{sohl2015deep}, DDPMs progressively corrupt the given data $x_0 \sim p(x_0)$ with noise over $T$ steps, following a variance schedule $\beta_t$ in the forward process.
The probability of the corrupted data can be modeled as follows:
\begin{equation}
\label{eq1: forward process}
q(x_{1:T}|x_0)=\prod_{t \geq 1}q(x_t|x_{t-1}), \indent
q(x_{x_t|x_{t - 1}}) = \mathcal{N}(x_t;\sqrt{1-\beta_t}x_{t-1},\beta_t I),
\end{equation}
where $I$ denotes the identity matrix. 
In the backword process, DDPMs iteratively remove noise from the corrupted data by estimating the mean $\mu_{\theta}(x_t, t)$ of the posterior as follows:
\begin{equation}
\label{eq2: backward process}
p_{\theta}(x_{t-1}|x_t)=\mathcal{N}(x_t-1;\mu_{\theta}(x_t, t),\sigma_t^2 I), \indent
\mu_{\theta}(x_t, t)=\frac{1}{\sqrt{\alpha_t}}(x_t-\frac{\beta_t}{\sqrt{1-\overline{\alpha}_t}}\epsilon_{\theta}(x_t, t)),
\end{equation}
where 
$\alpha_t$ represents the noise reduction factor at time step $t$ defined as $\alpha_t=1-\beta_t$, and $\overline{\alpha}_t$ is the cumulative value calculated by $\overline{\alpha}_t=\prod_{s=1}^t{\alpha_s}$. 
Note that $\epsilon_{\theta}(x_t, t)$ is typically trained as a U-Net structure to approximate the noise in the reverse process.

\subsubsection{Tensor networks (TNs)~\cite{cichocki2014tensor} for PEFT.}
\begin{figure}[t]
    \centering
    \includegraphics[width=0.95\textwidth]{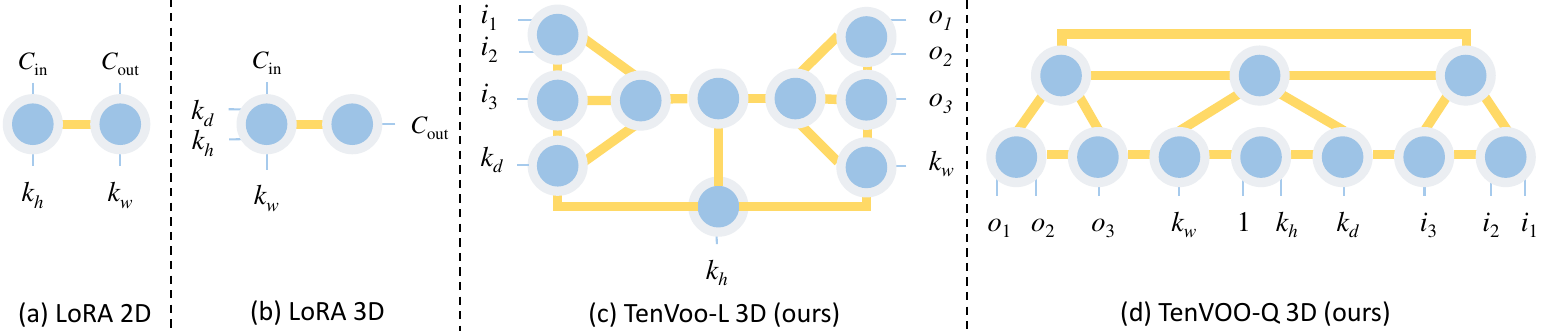}
    \caption{Structure of (a) LoRA for 2D convolution\protect\footnotemark[1]; (b) LoRA for 3D covolution\protect\footnotemark[2]; (c) TenVOO-L for 3D convolution and (d) TenVOO-Q for 3D convolution. The \textbf{\textcolor{yellow}{thick yellow}} connections represent the tunable rank of the TN.}
    \label{fig2:tns}
\end{figure}

TNs are a family of structural representations for high-dimensional tensors, where a tensor can be decomposed into a set of lower-dimensional core tensors interconnected through contraction operations. 
Following this framework, TNs can be visualized with graphs, in which core tensors correspond to nodes and index contractions are represented as edges. 
In the context of PEFT for example, the well-known LoRA \cite{hu2022lora} for 2D convolutional layers can be formulated as follows:
\begin{equation}
\label{eq3: lora2d}
\begin{gathered}
\Delta \mathcal{W} = reshape(B \times_{r} A, \left[ C_{out}, C_{in}, k_h, k_w \right]), \\
A \in \mathbb{R}^{r \times C_{in} \times k_h}, \indent
B \in \mathbb{R}^{C_{out} \times k_w \times r},
\end{gathered}
\end{equation}
where $\Delta \mathcal{W}$ is the low-rank update of the 2D convolutional weights, $r$ corresponds to the LoRA rank, and $C_{out}, C_{in}, k_h, k_w$ represent the number of output channels, input channels, and the kernel size in height and width, respectively. 
Note that the operation $\times_r$ represents the contraction between $A$ and $B$ along the index ``$r$''.
For the $\{o, i, h, w\}$-th entry of $\Delta \mathcal{W}$, this contraction can formulated with
$\Delta \mathcal{W} _ {o, i, h, w} = \sum _ {j=1} ^ {r} B_{o, w, j} A_{j, i, h}$.
Furthermore, it can be visualized by a graph as shown in Figure~\ref{fig2:tns}~(a).

As TNs can model complex internal dependencies through contractions among low-rank tensors using few trainable parameters, they have been widely adopted in PEFT, particularly for Transformer-based architectures \cite{jie2023fact,bershatsky2024lotr,chen2024quanta}.
Among these, QuanTA \cite{chen2024quanta} introduces a TN inspired by quantum circuits, facilitating high-rank representations through the contraction of a set of low-dimensional tensors, thereby preserving the representation capacity.

\subsection{Problem Formulation}
Recall that the objective of this work is PEFT for U-Net-based DDPMs in MRI image generation.
In a standard 3D convolutional layer with ignoring the activation function, we know that the output $\mathcal{Y}$ is given by
\begin{equation}
\label{eq4: conv3d}
\mathcal{Y} = \mathcal{W} * \mathcal{X} + b,
\end{equation}
where  $\mathcal{X}$ is the input feature map,  $\mathcal{W}\in\mathbb{R}^{C_{out} \times C_{in} \times k_d \times k_h \times k_w}$ is the weight tensor,  $*$ denotes the 3D convolution, and $b$ is the bias.
Since high-dimensional 3D-convolutional kernels $\mathcal{W}$ require substantial memory and computational resources, the main idea of the proposed method is thus to \emph{efficiently represent the difference in $\mathcal{W}$ during fine-tuning using TNs}, i.e., 
\begin{equation}
\label{eq5: tn for conv3d}
\mathcal{Y} = (\mathcal{W} + \Delta \mathcal{W}) * \mathcal{X} + b, \indent
\Delta \mathcal{W} = reshape(\mathcal{T}_{\theta} (t_1, ..., t_k), \left[ C_{out}, C_{in}, k_d, k_h, k_w \right]),
\end{equation}
where $\{ t_1, ..., t_k \}$ represents $k$ core tensors constituting the TN. 
With the updates $\Delta  \mathcal{W}$ parameterized by a TN, we only need to update those lower-dimensional core tensors during the fine-tuning phase.

\footnotetext[1]{\href{https://github.com/microsoft/LoRA/blob/c4593f060e6a368d7bb5af5273b8e42810cdef90/loralib/layers.py}{LoRA implementation for 2D convolutional layer}}
\footnotetext[2]{\href{https://github.com/KohakuBlueleaf/LyCORIS/blob/main/lycoris/modules/locon.py}{LoRA implementation for 3D convolutional layer (LoCon) from Lycoris}}

\subsection{Tensor Volumetric Operator}



In this work, we propose two fine-tuning methods for 3D convolutional layers, named TenVOO-L and TenVOO-Q, inspired by LoRA \cite{hu2022lora} and QuanTA \cite{chen2024quanta}.

TenVOO-L represents the kernels using a TN structure, as illustrated in Figure \ref{fig2:tns} (c).
Unlike the vanilla LoRA in Figure \ref{fig2:tns} (a), TenVOO-L first tensorizes $C_{in}=i_1 \times i_2 \times i_3$ and $C_{out}=o_1 \times o_2 \times o_3$, and then represents the weights via the TN. It is worth noting that the spatial dimensions are placed separately in TenVOO-L, enabling the model to capture spatial dependencies through tensor contraction. 
Moreover, when one convolutional kernel dimension is removed (e.g., $k_h$ in Figure \ref{fig2:tns}~(c)), our TenVOO-L degenerates into the 2D convolution.

TenVOO-Q, as shown in Figure \ref{fig2:tns}(d), is a direct extension of QuanTA, but with spatial dimensions explicitly assigned to different cores. Since QuanTA is renowned for maintaining high-rank representations while leveraging small-sized tensors as inputs, TenVOO-Q is expected to perform well for 3D convolutional layer with larger input or output channels, or larger kernel sizes.
The corresponding number of trainable parameters for TenVOO-L and TenVOO-Q, denoted respectively $\#P_{L}$ and $\#P_{Q}$ is given as
\begin{equation}
\label{eq6:P}
\begin{gathered}
\#P_{L} = (i_1 i_2 + o_1 o_2) r^2 + (i_3 + o_3 + k_d + k_h + k_w + 1) r^3 + 2r^4, \\
\#P_{Q} = (i_1 i_2 + o_1 o_2 + k_h) r^2 + (i_3 + o_3 + k_d + k_w) r^3 + 3r^4.
\end{gathered}
\end{equation}
We can see from Eqs.~\eqref{eq6:P} that the parameter scale using TenVOO is primarily determined by  $r$, which corresponds to the TN ranks, controling the trade-off between representation power and parameter efficiency in the model.

\noindent
\textbf{Linear layers follows QuanTA.}
Since our DDPM also includes linear layers, our TenVOO directly integrates QuanTA \cite{chen2024quanta} for fine-tuning. Following their approach, we tensorize both the input and output feature dimensions into third-order tensors and then construct the QuanTA network to model these layers. Note that the linear layers account for only about 10\% of the trainable parameters in our DDPM.

\subsection{Initialization Is Important for TenVOO.}
To ensure stability in training, PEFT typically requires that the weight updates in the adapted model remain at zero during initialization. 
Although the core tensors in TenVOO can follow the same routine, we empirically found that this routine negatively impacts the performance of fine-tuning in practice.
To address this, We adopt the initialization method of QuanTA for the convolutional layers, introducing a frozen copy of the trainable TN for initialization. In particular, given a trainable TN 
$\mathcal{T}_{\theta} (t_1, ..., t_k)$, 
we define its frozen copy 
$\mathcal{T}^{\includegraphics[height=0.5em]{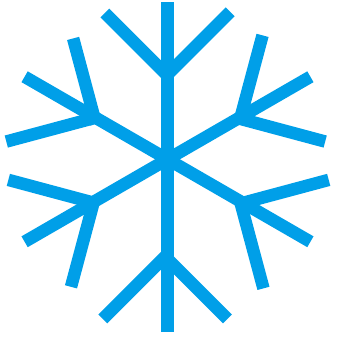}}$ 
at the beginning of training as:
\begin{equation}
\label{eq7: frozen copy}
\mathcal{T}^{\includegraphics[height=0.5em]{figures/snow.pdf}} = 
\mathcal{T}_{\theta} (t_1^{(0)}, ..., t_k^{(0)}),
\end{equation}
where $\{ t_1^{(0)}, ..., t_k^{(0)} \}$ are the initialized core tensors before training. 
The weight updates during fine-tuning is then modified as
\begin{equation}
\begin{gathered}
\label{eq8: updates}
\mathcal{Y} = (\mathcal{W} - \mathcal{T}^{\includegraphics[height=0.5em]{figures/snow.pdf}} + \Delta \mathcal{W}) * \mathcal{X} + b, \\
\Delta \mathcal{W} = reshape(\mathcal{T}_{\theta} (t_1, ..., t_k), \left[ C_{out}, C_{in}, k_d, k_h, k_w \right]).
\end{gathered}
\end{equation}
Here only core tensors $\{ t_1^{(0)}, ..., t_k^{(0)} \}$ in $\Delta \mathcal{W}$ are updated during fine-tuning. For linear layers, they follow similar initialization methods.

\section{Experiments and Results} \label{sec:exp}



\subsection{Dataset and experimental settings}
We use 59,830 T1-weighted MRI scans from UK Biobank (UKB) \cite{sudlow2015uk} to pre-train an unconditional U-Net-based DDPM structured as in Figure \ref{fig1:framework}, and 979, 545 and 327 T1-weighted scans from the Parkinson's Progression Markers Initiative (PPMI) \cite{marek2011parkinson}, Alzheimer’s Disease Neuroimaging Initiative (ADNI) \cite{jack2008alzheimer} databases, and BraTS2021 \cite{baid2021rsna} dataset for fine tuning, respectively.
For \textbf{UKB}, all scans are officially resampled to 1 mm$^{3}$, skull-stripped, registered to the MN152 space.
For \textbf{BraTS2021}, scans are also officially resampled to 1 mm$^{3}$ and skull stripped. We follow existing work \cite{dorjsembe2024conditional} to filter 327 images out of 1,251 training samples based on visual quality.  
For both \textbf{PPMI} and \textbf{ADNI}, the scans are resampled to 1 mm$^{3}$, skull stripped, bias corrected, tissue-segmented, registered to the MN152 space, and normalized, with all preprocessing steps performed using the CAT12 \cite{gaser2024cat} toolbox. 
Our usage has been approved by UKB.

For pre-training, we build our DDPM using MONAI \cite{cardoso2022monai} and train it on all 59,830 images from UKB. For fine-tuning, we adapt only the ResNet convolutional layers, the value and query projection layers in ResNet attention, and the time embedding and time projection layers in DDPM as our target modules.
We use 90\% of samples for training and the remaining 10\% for evaluation.
We pad and crop images from the UK Biobank, PPMI and ADNI to a size of $160 \times 224 \times 160$, and images from BraTS2021 to $192 \times 192 \times 144$, followed by resizing them to $128 \times 128 \times 96$. For training, all settings are trained using Mean Squared Error (MSE) loss, with learning rate of 0.00005, a batch size of 1 with 4 accumulation steps, optimized with the Adam optimizer \cite{kingma2014adam}. We set tensor rank as 4 and constrain our TenVOO to have the least number of trainable parameters.
For evaluation, we examine the generation quality using Fréchet Inception Distance (FID) \cite{heusel2017gans} and Maximum Mean Discrepancy (MMD) \cite{gretton2012kernel}. Since both metrics require an encoder to map images into a representational space, we follow existing work \cite{dorjsembe2024conditional,pinaya2022brain} and use Med3D \cite{chen2019med3d} as the encoder. We evaluate the spatial similarity using Multi-Scale Structural Similarity Index Measure (MS-SSIM) \cite{wang2003multiscale}.  
We implement existing PEFT methods for convolution, including LoRA \cite{hu2022lora}, LoKr \cite{edalati2022krona}, and LoHa \cite{hyeon2021fedpara}, as baselines, following the Lycoris \cite{yeh2024navigating}. 


\subsection{Results}

\begin{table}[t]
    \centering
    \caption{Evaluation of generation quality using FID$\downarrow$ and MMD$\downarrow$, along with structural similarity measured by MS-SSIM$\uparrow$. Here, Full-FT means full fine-tuning, MS-SSIM is denoted as ``MS'', and ``\#P'' represents the number of trainable parameters (in millions). All FID and MMD values are scaled by $10^{-2}$. The best values for each metric are highlighted in \textbf{bold}.}
    \label{tab1:quality}
    \renewcommand{\arraystretch}{1.2} 
    \begin{tabular}{lc ccc ccc ccc}
        \specialrule{1.2pt}{0pt}{0pt}
        \multirow{2}{*}{Model} & \multirow{2}{*}{\#P ($M$)} & \multicolumn{3}{c}{\textbf{ADNI}} & \multicolumn{3}{c}{\textbf{PPMI}} & \multicolumn{3}{c}{\textbf{BraTS2021}} \\
        \cline{3-5} \cline{6-8} \cline{9-11}
        & & FID & MMD & MS & FID & MMD & MS & FID & MMD & MS \\
        \specialrule{1.2pt}{0pt}{0pt}
        Real & -- & 0.016 & -- & 0.928 & 0.014 & -- & 0.934 & 0.004 & -- & 0.726 \\
        \hline
        Full-FT & 166.67 & 12.001 & 7.760 & 0.701 & 19.751 & 16.177 & 0.646 & 4.006 & 1.366 & 0.436 \\
        \hline
        LoRA \cite{hu2022lora} & 0.82 & 16.794 & 12.729 & 0.511 & \textbf{14.609} & \textbf{10.482} & 0.431 & 8.648 & 4.330 & 0.076 \\
        LoKr \cite{edalati2022krona} & 0.79 & 14.626 & 10.460 & 0.434 & 15.857 & 11.881 & 0.253 & 3.399 & 1.086 & 0.291 \\
        LoHa \cite{hyeon2021fedpara} & 0.82 & 17.031 & 13.166 & 0.490 & 14.677 & 10.529 & 0.426 & 8.759 & 4.452 & 0.084 \\
        \hline
        TenVOO-L & 0.60 & 17.349 & 13.331 & \textbf{0.663} & 16.605 & 12.524 & 0.668 & \textbf{3.168} & \textbf{0.908} & \textbf{0.581} \\
        TenVOO-Q & 0.58 & \textbf{13.475} & \textbf{9.502} & 0.504 & 19.585 & 15.700 & \textbf{0.756} & 3.234 & 0.934 & 0.575 \\
        \specialrule{1.2pt}{0pt}{0pt}
    \end{tabular}
\end{table}

\begin{table}[t]
    \centering
    \caption{Evaluation results for jointly fine-tuning. All settings of the table are the same as those in Table \ref{tab1:quality}.}
    \label{tab2:jointly}
    \renewcommand{\arraystretch}{1.2} 
    \begin{tabular}{lc ccc ccc ccc}
        \specialrule{1.2pt}{0pt}{0pt}
        \multirow{2}{*}{Model} & \multirow{2}{*}{\#P ($M$)} & \multicolumn{3}{c}{\textbf{ADNI}} & \multicolumn{3}{c}{\textbf{PPMI}} & \multicolumn{3}{c}{\textbf{BraTS2021}} \\
        \cline{3-5} \cline{6-8} \cline{9-11}
        & & FID & MMD & MS & FID & MMD & MS & FID & MMD & MS \\
        \specialrule{1.2pt}{0pt}{0pt}
        \hline        
        LoRA \cite{hu2022lora} & 47.59 & 13.856 & 9.471 & 0.743 & 21.950 & 18.833 & 0.607 & 4.444 & 1.499 & 0.519 \\
        LoKr \cite{edalati2022krona} & 47.55 & \textbf{11.594} & \textbf{7.359} & 0.211 & 20.472 & 17.081 & 0.494 & 5.355 & 2.087 & 0.211  \\
        LoHa \cite{hyeon2021fedpara} & 47.59 & 12.502 & 8.257 & 0.699 & 23.032 & 20.171 & 0.613 & 4.740 & 1.609 & 0.436 \\
        TenVOO-L & 47.36 & 12.153 & 7.899 & \textbf{0.804} & 11.862 & \textbf{7.539} & 0.771 & \textbf{1.256} & \textbf{0.190} & \textbf{0.715}\\
        TenVOO-Q & 47.34 & 12.676 & 8.756 & 0.507 & \textbf{11.857} & 7.594 & \textbf{0.837} & 2.594 & 0.737 & 0.462 \\
        \specialrule{1.2pt}{0pt}{0pt}
    \end{tabular}
\end{table}

\begin{figure}[t]
    \centering
    \includegraphics[width=0.75\textwidth]{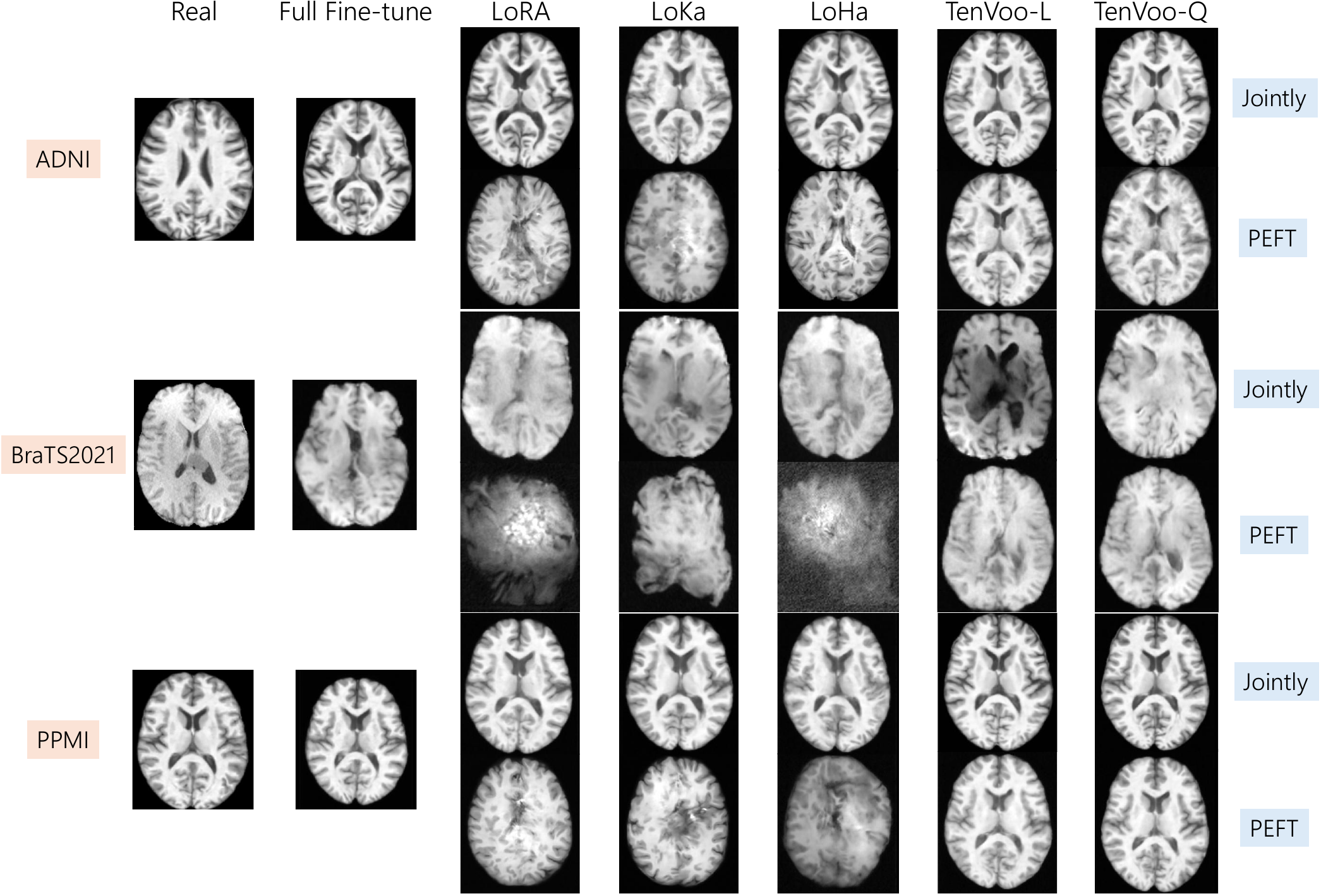}
    \caption{Visualization results. For each image pair, the top scan is generated by the jointly fine-tuned model, while the bottom scan is generated by the adapted model.}
    \label{fig3:vis}
\end{figure}

\begin{figure}[htpb]
    \centering
    \includegraphics[width=0.995 \textwidth]{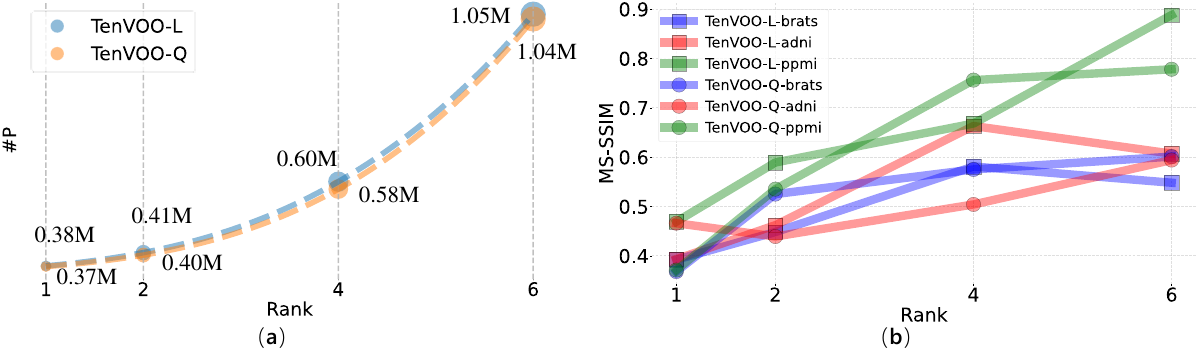}
    \caption{Results of the ablation study on rank. (a) illustrates how the number of parameters changes as the rank increases. (b) demonstrates the impact of rank on MS-SSIM performance.}
    \label{fig4:rank}
\end{figure}

Table \ref{tab1:quality} presents the results of both generation quality and structural similarity on the evaluation set. Our TenVOOs achieve competitive performance in terms of FID and MMD compared to the baselines while outperforming them on MS-SSIM. 
Notably, we observe that TenVOOs yield much lower FID and MMD scores on PPMI. We speculate that T1-weighted MRI scans for Parkinson's disease exhibit minimal structural variations \cite{burton2004cerebral,melzer2012grey}, making it challenging to adapt only the TNs in our method, especially those used for convolutions, which are designed to effectively capture spatial information, to represent them well.
Additionally, TenVOOs significantly improve performance on BraTS2021, which differs the most from the pre-training data in the UKB dataset.


For a more comprehensive comparison, we also present results from jointly fine-tuned models, as given in Table \ref{tab2:jointly}. In this setting, in addition to adapting the target modules with low-rank representations, other layers, such as up-sampling and down-sampling layers, skip connections, and others, are also updated using the standard full fine-tuning approach.
The overall visual results are illustrated in Figure \ref{fig3:vis}. 
We can see that images generated by jointly fine-tuned models retain the visual structure, while those from the baseline models exhibit varying degrees of structural distortion. In contrast, our TenVOOs effectively preserve the spatial integrity of brain MRI scans.

\noindent \textbf{Ablation study.} 
We conduct an ablation study on the tensor rank of our TenVOO. Specifically, we vary the rank across {1,2,4,6} and evaluate its impact on \#P and MS-SSIM. Figure \ref{fig4:rank}~(a) shows that our TenVOO can maintain a relatively low number of trainable parameters as the rank increases. Figure \ref{fig4:rank}~(b) demonstrates that our TenVOO effectively learns representations for spatial information, as indicated by the increasing MS-SSIM scores with higher ranks.


\section{Conclusion}

We introduced TenVOO, a novel parameter-efficient fine-tuning (PEFT) method for adapting 3D convolutional layers in U-Net-based DDPMs for MRI image generation. By leveraging tensor networks, TenVOO efficiently captures complex spatial dependencies while significantly reducing trainable parameters. Experimental results across multiple datasets demonstrate its effectiveness in improving structural similarity and generation quality, making it a promising approach for optimizing 3D convolutional models in medical imaging. Future work includes extending TenVOO to broader generative tasks and further refining its efficiency for large-scale applications.

\bibliographystyle{splncs04}

\begin{thebibliography}{10}
\providecommand{\url}[1]{\texttt{#1}}
\providecommand{\urlprefix}{URL }
\providecommand{\doi}[1]{https://doi.org/#1}

\bibitem{ali2022spot}
Ali, H., Murad, S., Shah, Z.: Spot the fake lungs: Generating synthetic medical images using neural diffusion models. In: AICS. pp. 32--39. Springer (2022)

\bibitem{baid2021rsna}
Baid, U., Ghodasara, S., Mohan, S., Bilello, M., Calabrese, E., Colak, E., Farahani, K., Kalpathy-Cramer, J., Kitamura, F.C., Pati, S., et~al.: The {RSNA-ASNR-MICCAI BraTS} 2021 benchmark on brain tumor segmentation and radiogenomic classification. arXiv preprint arXiv:2107.02314  (2021)

\bibitem{bershatsky2024lotr}
Bershatsky, D., Cherniuk, D., Daulbaev, T., Mikhalev, A., Oseledets, I.: {LoTR}: Low tensor rank weight adaptation. arXiv preprint arXiv:2402.01376  (2024)

\bibitem{burton2004cerebral}
Burton, E.J., McKeith, I.G., Burn, D.J., Williams, E.D., O’Brien, J.T.: Cerebral atrophy in {P}arkinson’s disease with and without dementia: a comparison with {A}lzheimer’s disease, dementia with lewy bodies and controls. Brain  \textbf{127}(4),  791--800 (2004)

\bibitem{cardoso2022monai}
Cardoso, M.J., Li, W., Brown, R., Ma, N., Kerfoot, E., Wang, Y., Murrey, B., Myronenko, A., Zhao, C., Yang, D., et~al.: {MONAI}: An open-source framework for deep learning in healthcare. arXiv preprint arXiv:2211.02701  (2022)

\bibitem{chavan2023one}
Chavan, A., Liu, Z., Gupta, D., Xing, E., Shen, Z.: One-for-all: Generalized {LoRA} for parameter-efficient fine-tuning. arXiv preprint arXiv:2306.07967  (2023)

\bibitem{chen2019med3d}
Chen, S., Ma, K., Zheng, Y.: Med3{D}: Transfer learning for 3{D} medical image analysis. arXiv preprint arXiv:1904.00625  (2019)

\bibitem{chen2024quanta}
Chen, Z., Dangovski, R., Loh, C., Dugan, O.M., Luo, D., Soljacic, M.: Quan{TA}: Efficient high-rank fine-tuning of {LLM}s with quantum-informed tensor adaptation. In: NeurIPS (2024), \url{https://openreview.net/forum?id=EfpZNpkrm2}

\bibitem{chung2022mr}
Chung, H., Lee, E.S., Ye, J.C.: {MR} image denoising and super-resolution using regularized reverse diffusion. IEEE Transactions on Medical Imaging  \textbf{42}(4),  922--934 (2022)

\bibitem{cichocki2014tensor}
Cichocki, A.: Tensor networks for big data analytics and large-scale optimization problems. arXiv preprint arXiv:1407.3124  (2014)

\bibitem{dorjsembe2024conditional}
Dorjsembe, Z., Pao, H.K., Odonchimed, S., Xiao, F.: Conditional diffusion models for semantic 3{D} brain {MRI} synthesis. IEEE Journal of Biomedical and Health Informatics  (2024)

\bibitem{edalati2022krona}
Edalati, A., Tahaei, M., Kobyzev, I., Nia, V.P., Clark, J.J., Rezagholizadeh, M.: Krona: Parameter efficient tuning with {Kronecker} adapter. arXiv preprint arXiv:2212.10650  (2022)

\bibitem{gaser2024cat}
Gaser, C., Dahnke, R., Thompson, P.M., Kurth, F., Luders, E., Initiative, A.D.N., et~al.: {CAT}: a computational anatomy toolbox for the analysis of structural {MRI} data. Gigascience  \textbf{13},  giae049 (2024)

\bibitem{gretton2012kernel}
Gretton, A., Borgwardt, K.M., Rasch, M.J., Sch{\"o}lkopf, B., Smola, A.: A kernel two-sample test. The Journal of Machine Learning Research  \textbf{13}(1),  723--773 (2012)

\bibitem{guo2023accelerating}
Guo, X., Yang, Y., Ye, C., Lu, S., Peng, B., Huang, H., Xiang, Y., Ma, T.: Accelerating diffusion models via pre-segmentation diffusion sampling for medical image segmentation. In: ISBI. pp.~1--5. IEEE (2023)

\bibitem{heusel2017gans}
Heusel, M., Ramsauer, H., Unterthiner, T., Nessler, B., Hochreiter, S.: {GAN}s trained by a two time-scale update rule converge to a local {Nash} equilibrium. In: NeurIPS (2017)

\bibitem{ho2020denoising}
Ho, J., Jain, A., Abbeel, P.: Denoising diffusion probabilistic models. In: NeurIPS. pp. 6840--6851 (2020)

\bibitem{houlsby2019parameter}
Houlsby, N., Giurgiu, A., Jastrzebski, S., Morrone, B., De~Laroussilhe, Q., Gesmundo, A., Attariyan, M., Gelly, S.: Parameter-efficient transfer learning for {NLP}. In: ICML. pp. 2790--2799. PMLR (2019)

\bibitem{hu2022lora}
Hu, E.J., Shen, Y., Wallis, P., Allen-Zhu, Z., Li, Y., Wang, S., Wang, L., Chen, W.: Lo{RA}: Low-rank adaptation of large language models. In: ICLR (2022), \url{https://openreview.net/forum?id=nZeVKeeFYf9}

\bibitem{hyeon2021fedpara}
Hyeon-Woo, N., Ye-Bin, M., Oh, T.H.: Fedpara: Low-rank hadamard product for communication-efficient federated learning. arXiv preprint arXiv:2108.06098  (2021)

\bibitem{jack2008alzheimer}
Jack~Jr, C.R., Bernstein, M.A., Fox, N.C., Thompson, P., Alexander, G., Harvey, D., Borowski, B., Britson, P.J., Whitwell, J.L., Ward, C., et~al.: The {A}lzheimer's disease neuroimaging initiative {(ADNI)}: {MRI} methods. Journal of Magnetic Resonance Imaging  \textbf{27}(4),  685--691 (2008)

\bibitem{jie2023fact}
Jie, S., Deng, Z.H.: Fact: Factor-tuning for lightweight adaptation on vision transformer. In: AAAI. vol.~37, pp. 1060--1068 (2023)

\bibitem{kingma2014adam}
Kingma, D.P., Ba, J.: Adam: A method for stochastic optimization. arXiv preprint arXiv:1412.6980  (2014)

\bibitem{marek2011parkinson}
Marek, K., Jennings, D., Lasch, S., Siderowf, A., Tanner, C., Simuni, T., Coffey, C., Kieburtz, K., Flagg, E., Chowdhury, S., et~al.: The {P}arkinson progression marker initiative {(PPMI)}. Progress in Neurobiology  \textbf{95}(4),  629--635 (2011)

\bibitem{melzer2012grey}
Melzer, T.R., Watts, R., MacAskill, M.R., Pitcher, T.L., Livingston, L., Keenan, R.J., Dalrymple-Alford, J.C., Anderson, T.J.: Grey matter atrophy in cognitively impaired {P}arkinson's disease. Journal of Neurology, Neurosurgery \& Psychiatry  \textbf{83}(2),  188--194 (2012)

\bibitem{packhauser2023generation}
Packh{\"a}user, K., Folle, L., Thamm, F., Maier, A.: Generation of anonymous chest radiographs using latent diffusion models for training thoracic abnormality classification systems. In: ISBI. pp.~1--5. IEEE (2023)

\bibitem{pinaya2022brain}
Pinaya, W.H., Tudosiu, P.D., Dafflon, J., Da~Costa, P.F., Fernandez, V., Nachev, P., Ourselin, S., Cardoso, M.J.: Brain imaging generation with latent diffusion models. In: MICCAI Workshop. pp. 117--126. Springer (2022)

\bibitem{ronneberger2015u}
Ronneberger, O., Fischer, P., Brox, T.: U-net: Convolutional networks for biomedical image segmentation. In: MICCAI. pp. 234--241. Springer (2015)

\bibitem{sohl2015deep}
Sohl-Dickstein, J., Weiss, E., Maheswaranathan, N., Ganguli, S.: Deep unsupervised learning using nonequilibrium thermodynamics. In: ICML. pp. 2256--2265. PMLR (2015)

\bibitem{sudlow2015uk}
Sudlow, C., Gallacher, J., Allen, N., Beral, V., Burton, P., Danesh, J., Downey, P., Elliott, P., Green, J., Landray, M., et~al.: {UK} biobank: an open access resource for identifying the causes of a wide range of complex diseases of middle and old age. PLoS medicine  \textbf{12}(3),  e1001779 (2015)

\bibitem{wang2003multiscale}
Wang, Z., Simoncelli, E.P., Bovik, A.C.: Multiscale structural similarity for image quality assessment. In: Asilomar Conference on Signals, Systems \& Computers. vol.~2, pp. 1398--1402. IEEE (2003)

\bibitem{wu2024medsegdiff}
Wu, J., Fu, R., Fang, H., Zhang, Y., Yang, Y., Xiong, H., Liu, H., Xu, Y.: Medsegdiff: Medical image segmentation with diffusion probabilistic model. In: MIDL. pp. 1623--1639. PMLR (2024)

\bibitem{yeh2024navigating}
Yeh, S.Y., Hsieh, Y.G., Gao, Z., Yang, B.B., Oh, G., Gong, Y.: Navigating text-to-image customization: From ly{CORIS} fine-tuning to model evaluation. In: ICLR (2024), \url{https://openreview.net/forum?id=wfzXa8e783}

\bibitem{zaken2021bitfit}
Zaken, E.B., Ravfogel, S., Goldberg, Y.: Bitfit: Simple parameter-efficient fine-tuning for transformer-based masked language-models. arXiv preprint arXiv:2106.10199  (2021)

\end{thebibliography}

\end{document}